\documentclass[11pt,paper,letterpaper]{JHEP3}
\usepackage{epsfig}
\usepackage{psfrag}
\usepackage{amsmath}
\newcommand{\bq}{\begin{eqnarray}}
\newcommand{\eq}{\end{eqnarray}}
\newcommand{\bqs}{\begin{eqnarray*}}
\newcommand{\eqs}{\end{eqnarray*}}

\def\f{\phi}
\def\g{\gamma}

\def\th{\theta}
\def\r{\rho}                                   
\def\s{\sigma}

\def\RN{Reissner-Nordstr\"om}
\title{The Petrov type of the BMPV metric}
\author{Pieter-Jan De Smet\\
C.~N. Yang Institute of Theoretical Physics\\
State University of New York\\
Stony Brook, NY 11794-3840, USA\\
E-mail: {\tt Pieterj@insti.physics.sunysb.edu} }
\preprint{YITP-SB-03-66 \\{\tt gr-qc/0401033}}
\abstract{We show that the BMPV metric has Petrov type 22. This means that
the BMPV metric is less algebraically special than the five-dimensional Schwarzschild
metric, which has Petrov type \underline{22}.}
\keywords{Classical Theories of Gravity, Black Holes}
\begin{document}
\section{Introduction}
In this article, we calculate the Petrov type of the BMPV metric~\cite{BMPV}, which is the 
metric of an extremal, charged, rotating black hole in minimal five-dimensional
supergravity~\cite{Cremmer}\footnote{See ref.~\cite{Reall} for the derivation of all
supersymmetric solutions in five-dimensional supergravity and~\cite{Townsend} for a 
discussion of general features of black holes in five dimensions.}. 
It turns out that it has Petrov type 22. To place this result in 
context, we remark that the five-dimensional Schwarzschild metric, the 
five-dimensional \RN\ metric and the five dimensional Myers-Perry metric~\cite{MP} 
all have Petrov type~\underline{22}. Therefore, adding electric charge \textit{or}
rotation to the five-dimensional black hole does not change its Petrov type. However,
adding both charge \textit{and} rotation makes the metric less algebraically special.
This result is in stark contrast with the behavior of the Petrov type of the analogous
four-dimensional metrics. Adding electric charge or rotation or both to the static
four-dimensional black hole does not change its Petrov type: the Schwarzschild,
\RN, Kerr and Kerr-Newman metric all have Petrov type~$D$.

This article is organized as follows. In Section~\ref{s:Petrov4}, 
we give a short
review of the four-dimensional Petrov classification. In Section~\ref{s:KN}, we calculate the 
Petrov type of the Kerr-Newman metric. In Section~\ref{s:Petrov5}, we 
review the five-dimensional Petrov classification. In Section~\ref{s:BMPV}, we 
calculate the Petrov type of the BMPV metric. We conclude in Section~\ref{s:conc}.
\section{Review of the Petrov classification in four dimensions}\label{s:Petrov4}
The Petrov classification in four dimensions is well-known. However, the Petrov 
classification in five dimensions, which we review in Section~\ref{s:Petrov5}, 
is less known. Therefore, we give a brief review of  the four-dimensional
Petrov classification to make the similarity with the five-dimensional 
Petrov classification clear. 

The Petrov classification in four dimensions is most easily discussed using 
two-component spinors~\cite{Penrose}. The Weyl spinor $C_{ABCD}$ is the 
spinor translation\footnote{We use the following conventions: $i,j,k$ and $l$ 
are vector indices and $A,B,C,D$ are spinor indices.
$\sigma_0 = 1$ and $\sigma_1, \sigma_2, \sigma_3$ are the Pauli matrices. 
The matrices $\bar{\sigma}_i$ are defined by $\bar{\sigma}_0 =\sigma_0$ and
 $\bar{\sigma}_i =-\sigma_i$ for $i = 1,2,3$.  Furthermore, we define
$$ \sigma^{ij}_{AB} = \epsilon_{AC} \sigma^{i C \dot{C} } \bar{\sigma}^j_{\dot{C}B}\ .$$} of the Weyl tensor $C_{ijkl}$
$$C_{ABCD} = \frac{1}{4} C_{ijkl} \sigma^{ij}_{AB} \sigma^{kl}_{CD}\ .$$
It is completely symmetric. 
The Petrov type of a Weyl tensor is given by the factorization properties
of the associated Weyl polynomial $W = C_{ABCD}\ x^A x^B x^C x^D$.
This polynomial is homogeneous of degree four in two variables. Therefore,
it can always be factorized as
\begin{equation}\label{factorsD4}
C_{ABCD} x^A x^B x^C x^D = (\alpha_A x^A)(\beta_B x^B) (\gamma_C x^C)(\delta_D x^D)\ .
\end{equation} 
In this way, we obtain six different Petrov types, see Figure~\ref{fig:D4}. 
\FIGURE[h]{
\begin{psfrags}
\psfrag{1}[][]{1111}
\psfrag{2}[][]{\underline{11}11}
\psfrag{3}[][]{\underline{11}\ \underline{11}}
\psfrag{4}[][]{\underline{111}1}
\psfrag{5}[][]{\underline{1111}}
\psfrag{6}[][]{$C_{ijkl} = 0$}
\epsfig{file=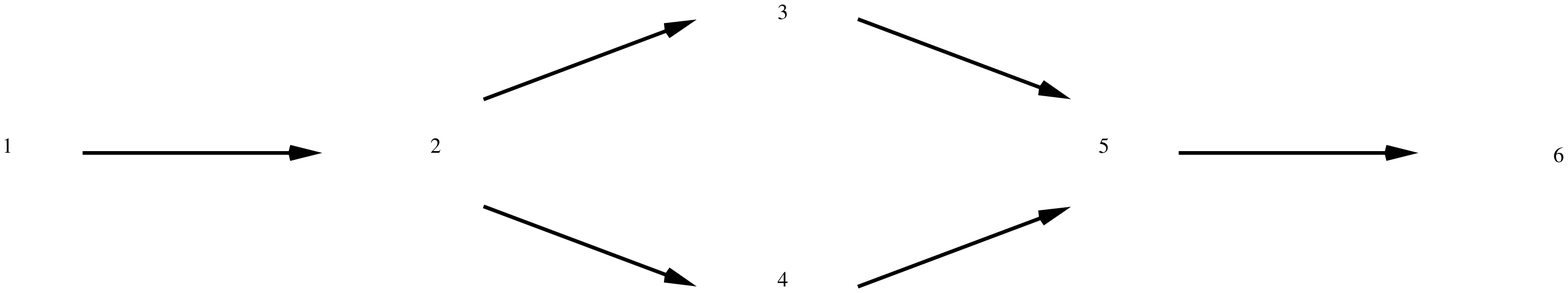,width=10cm,height=3cm}
\end{psfrags} 
\caption{The Penrose diagram of the six different Petrov types in four 
dimensions. We use underbars to denote how many factors in the factorization~(\ref{factorsD4})
coincide; the metric has Petrov type \underline{11}11, \underline{111}1 or \underline{1111}
if two, three, or respectively all factors coincide. The metric has Petrov type 
\underline{11}\ \underline{11} if two different sets of factors coincide. 
This case is usually called Petrov type~$D$. Petrov type 1111 is algebraically general, 
in this case all factors are different.
}
\label{fig:D4}
}

In the next section, we will also use the Maxwell spinor $F_{AB}$, which
is the spinor translation of the electromagnetic field strength
$F_{AB} = \frac{1}{2} \sigma^{ij}_{AB}  F_{ij}$.
With this bispinor, we associate the Maxwell polynomial
$M= F_{AB} x^A x^B$.
\section{The Kerr-Newman metric has Petrov type $D$}\label{s:KN}
We use the following tetrad for the Kerr-Newman metric~\cite{KN}
$$
e^1 = \dfrac{R}{\r} \left(dt - a \sin^2\th d\f\right), \
e^2 =  \dfrac{\sin\th}{ \r} \left(\left(r^2 + a^2\right) d\f - a dt\right),\
e^3 =\dfrac{\r}{R} dr,\
e^4 = \r d\th,
$$
where $\r^2 = r^2 + a^2 \cos^2\th$ and $R^2 = r^2 - 2 m r + a^2 + Q^2$. This 
tetrad, together with the gauge potential 
$$ A = \frac{2 Q r}{\r R} e^1,$$
is a  solution of the Einstein-Maxwell equations. A short calculation
gives the Maxwell polynomial
$$M = \frac{ 2 i Q}{\bar z^2} \left( (x^1)^2 +(x^2)^2\right)\quad, \mbox{ where }\quad
z = r + i a \cos\theta.$$
It turns out that the Weyl polynomial is proportional to the square of the
Maxwell polynomial
$$W = \frac{3 \bar z (Q^2 - m z)}{2 Q^2 z} \ M^2.$$
The polynomial can be factorized as $\sim (x^1 + i x^2)^2 (x^1 - i x^2)^2$. Therefore,
the Kerr-Newman metric has Petrov type $D$.
\section{Review of the Petrov classification in five dimensions}\label{s:Petrov5}
As in four dimensions, the Petrov classification in five dimensions is a 
classification of the Weyl tensor.
We only give a brief review of this classification, a longer discussion can 
be found in ref.~\cite{0206106}. For a review of the algebraic classification
of the Ricci tensor in five dimensions, see ref.~\cite{0312064}.

In five dimensions, it is again
natural to use spinors to discuss the Petrov 
classification\footnote{For the Petrov classification in five and higher dimensions using tensor methods:
see ref.~\cite{0401ccc}. It would be good to make a comparison between the five-dimensional
Petrov classification using spinors, as discussed in Section~4, and the (more complicated) Petrov classification
using tensors.}.
As in four dimensions, we define the Weyl spinor
$C_{abcd}$, which 
is the spinorial translation of the Weyl tensor $C_{ijkl}$,
$$C_{abcd} = (\g_{ij})_{ab} (\g_{kl})_{cd}C^{ijkl}.$$
Here, $\g_{ij} = \frac{1}{2} [\g_i,\g_j ]$, where $\g_i$ are
the $\g$-matrices in five dimensions\footnote{In Section~5, we will use the following representation
$\g_1 = i \s_1 \otimes 1$, $\g_2 = \s_2\otimes 1$, $\g_3 = \s_3\otimes \s_1$,
$\g_4 = \s_3 \otimes \s_2$ and $\g_5 = \s_3\otimes \s_3$.}. 
 The Weyl spinor is 
symmetric in all its indices.
The Weyl polynomial $W$ is a homogeneous polynomial of degree four in four variables:
$$
W = C_{abcd} x^a x^b x^c x^d\ .
$$
As in four dimensions, the Petrov type of a given Weyl tensor is the number and
multiplicity of the irreducible
factors of its corresponding Weyl polynomial $W$. 
In this way, we obtain 12 different
Petrov types, which are depicted in Figure~\ref{fig:Ptypes}.
\FIGURE[h]{
\epsfxsize=10cm
\epsfysize=5cm
\begin{psfrags}
\psfrag{1}[][]{4}
\psfrag{2}[][]{31}
\psfrag{3}[][]{22}
\psfrag{4}[][]{211}
\psfrag{5}[][]{\underline{22}}
\psfrag{6}[][]{2\underline{11}}
\psfrag{7}[][]{1111}
\psfrag{8}[][]{\underline{11} \underline{11}}
\psfrag{9}[][]{11\underline{11}}
\psfrag{10}[][]{$C_{ijkl} = 0$}
\psfrag{11}[][]{\underline{1111}}
\psfrag{12}[][]{1\underline{111}}
\epsfbox{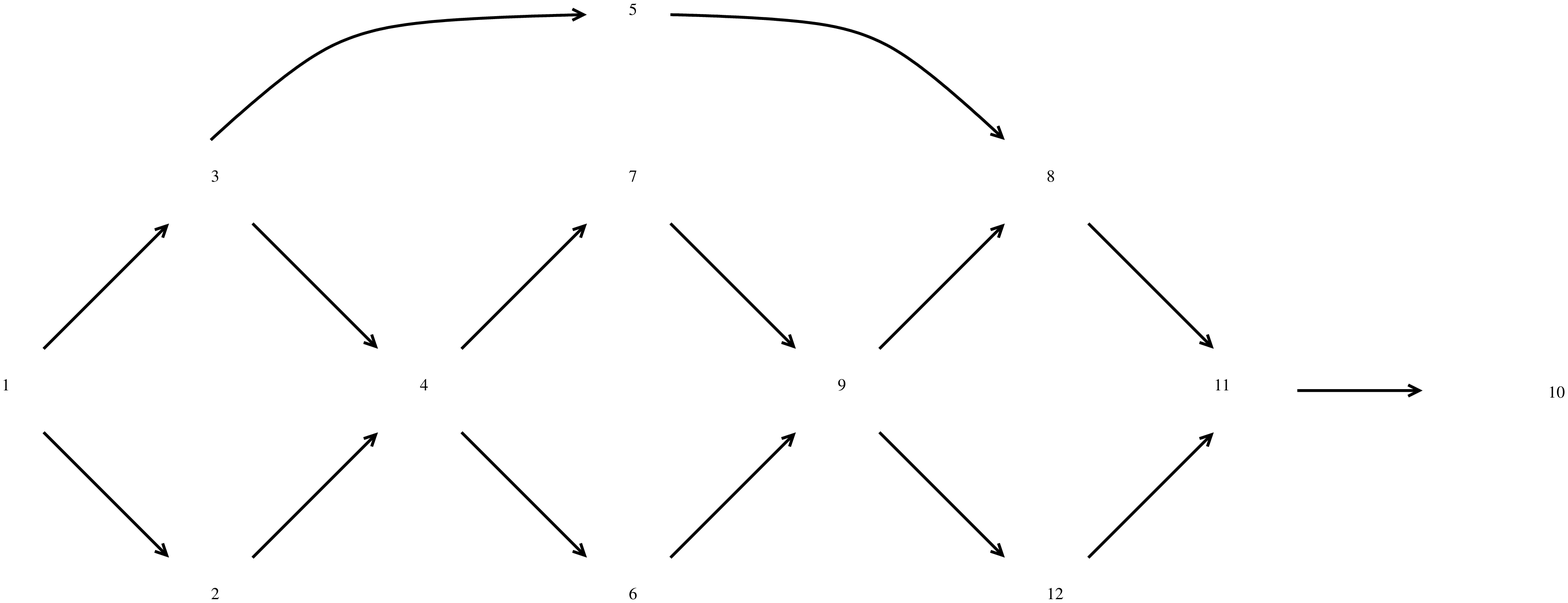}
\end{psfrags} 
\caption{The twelve different Petrov types in five dimensions.
We use the following notation. The number denotes the degree of the
irreducible factors and underbars denote the multiplicities. For example, a Weyl
polynomial which can be factorized into two different factors, each having
degree~2, has Petrov type~22. If the two factors of degree~2 are the same, the 
Petrov type is~\underline{22}.}
\label{fig:Ptypes}
}
We also define the Maxwell spinor 
$F_{ab} = (\gamma_{ij})_{ab} F^{ij}$ and the Maxwell polynomial 
$M = F_{ab} x^a x^b$.
\section{The BMPV metric has Petrov type 22}\label{s:BMPV}
We use the following tetrad for the BMPV metric~\cite{BMPV}
\begin{align*}
e^1 &= f(r)^{-1}\left[dt+\frac{\mu l}{r^2}(\sin^2\theta d\phi-\cos^2\theta d\psi)\right],\quad
e^2 = f(r)^{1/2} r \sin\theta d\phi,\\
e^3 &= f(r)^{1/2}r \cos\theta d\psi,\quad
e^4 = f(r)^{1/2} dr,\quad
e^5 = f(r)^{1/2} r d\theta,
\end{align*}
with $f(r) =  1+\frac{\mu}{r^2}$. The gauge potential is
$A = \sqrt{3 }\ e^1$.
A short calculation gives the Maxwell polynomial
\bqs
M = \frac{4 \sqrt{3} \mu}{ r^2 f(r)^2}& \big[ & r f(r)^{1/2} \left( (x^1)^2 -(x^2)^2 + (x^3)^2 - (x^4)^2 \right) \\
&& +l \sin\theta \left( (x^1 + x^2)^2 + (x^3 + x^4)^2 \right) - 2 i l \cos\theta \left( x^1 + x^2\right)
\left( x^3 + x^4\right) \big]\ .
\eqs
The Weyl polynomial is given by $ W = M       N$, with 
$$ N = \frac{1}{4} M - \frac{4 \sqrt{3}}{r f(r)^{3/2} } \left[ (x^1)^2 - (x^2)^2 +(x^3)^2 - (x^4)^2 \right].$$ 
Because the Weyl polynomial is the product of two factors of degree two, the Petrov type 
of the BMPV metric is~22. Hence, we see that the BMPV metric is less algebraically 
special than the Schwarzschild metric, which has Petrov type~\underline{22}, see~\cite{0206106}. On the
other hand, the five-dimensional Reissner-Nordstr\"om metric and the five-dimensional
Myers-Perry metric~\cite{MP} have both Petrov type~\underline{22} (the Petrov type
of the latter metric has been calculated in ref.~\cite{0312021}). This result is in 
contrast with the four-dimensional case, where adding charge or rotation or both to the
Schwarzschild metric does not change its Petrov type, see Section~3.
However, as in four dimensions, it is still true that the Maxwell polynomial
divides the Weyl polynomial. A physical reason for this is, as far as I know, not known. 
\section{Conclusions and topics for further research}\label{s:conc}
In this paper, we have shown that the Petrov type of the BMPV metric is~22. This 
means that the metric of the charged rotating black hole in five dimensional
minimal supergravity is less algebraically special that the 
five-dimensional Schwarzschild metric. This result is in contrast with
the four-dimensional case, where the Schwarzschild metric has the 
same Petrov type as its charged and rotating cousins.
Some topics for further research are the following.
\begin{itemize}
\item
The BMPV black hole is extremal: its electrical charge is equal to its
mass and its two angular momenta are equal. It would be good to calculate the Petrov type 
of its non-extremal generalizations. These metrics were derived in~\cite{Cvetic}, 
explicit expressions for some special cases can be found in~\cite{Herdeiro}.
\item
The action of five-dimensional minimal supergravity contains
a Chern-Simons term $A \wedge F \wedge F$ with a particular coefficient fixed by
supersymmetry. The metric of a charged rotating
black hole in five dimensions is not known
when this Chern-Simons term has an arbitrary (or even zero) coefficient. 
One might look for these metrics
within the class of metrics of Petrov type~22. As a further simplification,
one might even try to assume that the Maxwell polynomial is a factor 
of the Weyl polynomial -- as is the case for the BMPV metric.
\end{itemize}
\section*{Acknowledgments}
This work has been supported in part by the NSF grant PHY-0098527.
I would like to thank C.A.R. Herdeiro for useful discussions.


\begin{thebibliography}{99}
\bibitem{BMPV}
J.~C.~Breckenridge, R.~C.~Myers, A.~W.~Peet and C.~Vafa,
``D-branes and spinning black holes,''
Phys.\ Lett.\ B {\bf 391} (1997) 93
[arXiv:hep-th/9602065].
\bibitem{Cremmer}
E.~Cremmer,
``Supergravities In 5 Dimensions,'' in 
Superspace And Supergravity. Proceedings, Nuffield Workshop, Cambridge, UK, 1980,
eds. S.~W.~Hawking and M.~Ro\v{c}ek. 
\bibitem{Reall}
J.~P.~Gauntlett, J.~B.~Gutowski, C.~M.~Hull, S.~Pakis and H.~S.~Reall,
``All supersymmetric solutions of minimal supergravity in five dimensions,''
Class.\ Quant.\ Grav.\  {\bf 20}, 4587 (2003)
[arXiv:hep-th/0209114].
\bibitem{Townsend}
J.~P.~Gauntlett, R.~C.~Myers and P.~K.~Townsend,
``Black holes of D = 5 supergravity,''
Class.\ Quant.\ Grav.\  {\bf 16}, 1 (1999)
[arXiv:hep-th/9810204].
\bibitem{MP}
R.~C.~Myers and M.~J.~Perry,
``Black Holes In Higher Dimensional Space-Times,''
Annals Phys.\  {\bf 172}, 304 (1986).
\bibitem{Penrose}
R.~Penrose, 
``A Spinor Approach to General Relativity,''
Annals Phys.\  {\bf   10}, 171-201 (1960).   
\bibitem{KN}
E.~T.~Newman et al., ``Metric of a Rotating, Charged Mass'' 
J. Math. Phys. {\bf 6}, 918-919, (1965).          
\bibitem{0206106}
P.~J.~De Smet,
``Black holes on cylinders are not algebraically special,''
Class.\ Quant.\ Grav.\  {\bf 19}, 4877 (2002)
[arXiv:hep-th/0206106].
\bibitem{0312064}
M.~J.~Reboucas, J.~Santos and A.~F.~F.~Teixeira,
``Classification of Energy Momentum Tensors in $n \geq 5$ Dimensional Space-times: a Review,''
arXiv:gr-qc/0312064.
\bibitem{0401ccc}
A.~Coley, R.~Milson, V.~Pravda and A.~Pravdov,
``Classification of the Weyl Tensor in Higher-Dimensions,''
arXiv:gr-qc/0401008,
``Alignment and algebraically special tensors in Lorentzian geometry,''
arXiv:gr-qc/0401010.
\bibitem{0312021}
P.~J.~De Smet,
``The Petrov type of the five-dimensional Myers-Perry metric,''
arXiv:gr-qc/0312021.
\bibitem{Cvetic}
M.~Cvetic and D.~Youm,
``General Rotating Five Dimensional Black Holes of Toroidally Compactified Heterotic String,''
Nucl.\ Phys.\ B {\bf 476}, 118 (1996)
[arXiv:hep-th/9603100].
\bibitem{Herdeiro}
C.~A.~R.~Herdeiro,
``The Kerr-Newman-Goedel black hole,''
Class.\ Quant.\ Grav.\  {\bf 20}, 4891 (2003)
[arXiv:hep-th/0307194].
\end{thebibliography}
\end{document}